# Detecting Emerging Technologies in Artificial Intelligence Scientific Ecosystem Using an Indicator-based Model

**Ali Ghaemmaghami[1], Andrea Schiffauerova[1,*] and Ashkan Ebadi[1,2]**

[1] Concordia University, Concordia Institute for Information Systems Engineering, Montreal, QC H3G 2W1, Canada

[2] National Research Council Canada, Montreal, QC H3T 2B2, Canada

* andrea@encs.concordia.ca

**Abstract:**

Early identification of emergent topics is of eminent importance due to their potential impacts on society. There are many methods for detecting emerging terms and topics, all with advantages and drawbacks. However, there is no consensus about the attributes and indicators of emergence. In this study, we evaluate emerging topic detection in the field of artificial intelligence using a new method to evaluate emergence. We also introduce two new attributes of collaboration and technological impact which can help us use both paper and patent information simultaneously. Our results confirm that the proposed new method can successfully identify the emerging topics in the period of the study. Moreover, this new method can provide us with the score of each attribute and a final emergence score, which enable us to rank the emerging topics with their emergence scores and each attribute score.

## 1. Introduction

Emerging technologies have the potential to alter not just the technological paradigms on which traditional industries rely or to generate entirely new industries (Day and Schoemaker 2000; Porter et al. 2002) but also to alter existing socio-economic structures and production practices (Adner and Snow 2010; Rotolo et al. 2015; Zhou et al. 2019). Early and accurate detection of emerging technologies can provide decision-makers with knowledge, intelligence, and opportunities, from research and development (R&D) departments of different institutions to national policy-making organizations and innovation administrations (Jang et al. 2021; S. Xu et al. 2021; Zhou et al. 2021). Especially in more recent years, the speed of technological changes is so rapid; therefore, fast detection of emerging technologies can be more valuable. For instance, some popular cutting-edge technologies such as cloud computing, mobile computing, the Internet of Things (IoT), the Internet of Services, data collection, big data analytics, Artificial Intelligence (AI), augmented reality, 3D printing, and other technologies developed and adopted quickly (Zamani et al. 2022). As the amount of data available to us is growing amazingly fast, applications of this data are growing as well. One of the applications of this huge data can be detecting emerging technologies or topics through data and with minimum intervention of experts.

The trend on the topic of emerging technologies is upward in recent years. We defined the search with the topic of “emerging technologies” in the Web of Science (WOS) database, and the result is shown in Figure 1. The topic of emerging technologies can be considered a hot topic in recent years, and also the subtopics around it such as emerging technologies detection.

There are different definitions for emerging technologies and methods for emerging technology detection. From lexical-based approaches (Joung and Kim 2017; Weismayer and Pezenka 2017; Wu and Leu 2014), bibliometric approaches (Daim et al. 2006; Kim and Bae 2017; Mejia and Kajikawa 2020), and indicator-based approaches (Abercrombie et al. 2012; Bengisu 2003; H. Xu et al. 2021) to more complex methods such as machine learning methods (Choi et al. 2021; S. Xu et al. 2021; Zhou et al. 2021) and hybrid methods (Ávila-Robinson and Miyazaki 2013; Carley et al. 2017; Q. Wang 2018). Most of these approaches use patents or publications as their source of emerging technologies. Based on their approach and their focus on different aspects, researchers tried to choose between patents and papers as their source of data. Both patents and papers can provide information about emerging technologies but at different times and levels. Ávila-Robinson and Miyazaki (2013) deployed both sources to capture the cycle of emerging technologies. However, very few tried to include both in the process of emerging technology detection. Mejia and Kajikawa (2020) were one of the few that evaluated the emerging topics in both science and technology using paper and patent databases simultaneously. We believe that deploying both of these sources in a method might provide us with the opportunity to use the maximum of all the useful data to detect emerging technologies in a meaningful way.

Many of the emerging technology detection methods rely on attributes of emergence to identify emerging technologies. There is not a consensus about the attributes of emergence. Rotolo et al. (2015) considered radical novelty, relatively fast growth, coherence, prominent impact, and uncertainty and ambiguity as attributes of emergence. Q. Wang (2018) took novelty, relatively fast growth, coherence, and scientific impact as attributes of emergent research topics. Carley et al. (2017) also used novelty, growth, persistence, and community as attributes of emergence in their method. However, there is still room for improvement and adding new attributes of emergence.

In the process of detecting emerging technologies, one of the essential steps is providing terms or topics. This can be done through a process called automatic keyword extraction (AKE) (Nasar et al. 2019). We use the novel method of Bidirectional Encoder Representations from Transformers (BERT) proposed by Devlin et al. (2019) as our keyword extraction method. We focus on the AI domain as a fast-growing subject to detect emerging technologies in that domain by using the WOS database.

We propose a new method to detect emergence by proposing two main attributes of collaboration and technological impact. The collaboration attribute can detect how well different actors can co-operate and work on a topic without using some complex methods. With the usage of technological impact as an attribute, we can maintain the data of patents while mainly focusing on papers as our source of data and overcome the problem of using only one source as our source of data. Additionally, we propose some metrics and sub-attributes that can solve some problems of previous methods in detecting hot-topics instead of emergent technologies. Moreover, with our proposed method, the attributes of emergence can directly and easily translate into unambiguous indicators to measure the degree of the emergence of a term or topic.

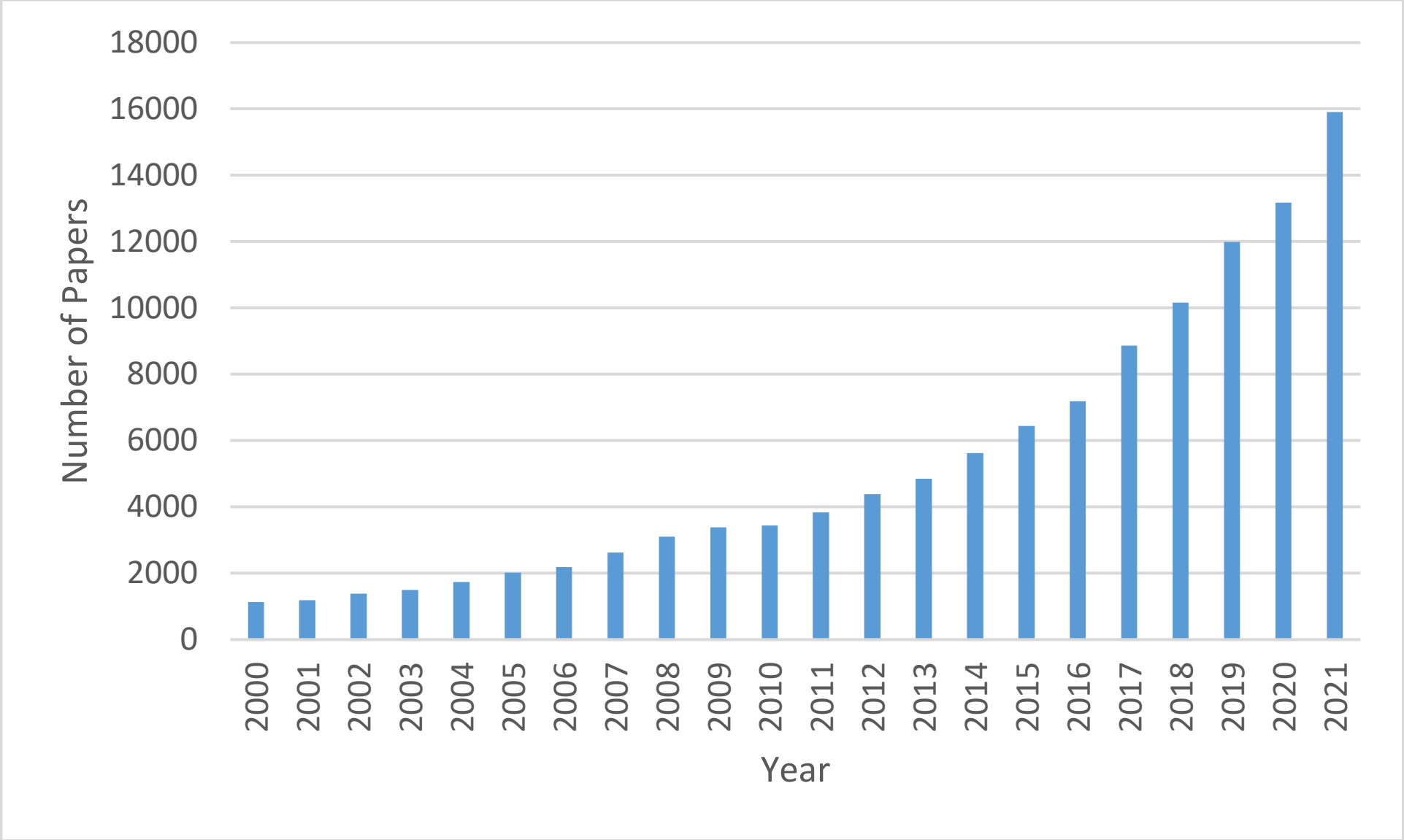

**Figure 1.** The trend of the documents with the topic of emerging technologies in the WOS database

## 2. Literature review

### 2.1. Definitions of Emergent Technologies

There are various definitions for the identical idea of emerging technologies and depending on the definition, different methods for detecting emerging technologies are applicable. We mainly focus on the following definition that utilizes defined attributes to detect emerging technologies. Cozzens et al. (2010) defined emerging technologies as those with characteristics such as rapid growth, newness, untapped market potential, and a high-technology base. Rotolo et al. (2015) believed that there are five attributes for considering a concept to be an emergence of novel technology: radical novelty, relatively fast growth, coherence, prominent impact, and uncertainty and ambiguity. Based on different circumstances, the definition can be altered; for instance, Q. Wang (2018) defined emergence for research topics by replacing scientific impact instead of prominent impact. Based on these definitions, many articles have been developed to design approaches to detect emergent technologies.

### 2.2. Approaches to Detect Emerging Technologies

Several approaches have been so far proposed for detecting emerging technologies. Rotolo et al. (2015) grouped them into five classes: 1. indicators and trend analysis 2. citation analysis 3. co-word analysis 4. overlay mapping, and 5. hybrid approaches. More recent papers included machine learning and excluded some obsolete methods such as the overlay mapping method from their categorization. S. Xu et al. (2021) grouped methods of emergence detection into three groups: citation-based approaches, lexical-based approaches, and machine learning approaches. We update the categorization with the latest approaches.

Based on the used methodology, we classify emerging detection approaches into five different groups: 1. Lexical-based approaches 2. Bibliometric-based approaches 3. Indicator-based approaches 4. Machine learning approaches and 5. Hybrid approaches. We will briefly review the most important works of each group.

### 2.2.1. Lexical-based approaches

Wu and Leu (2014) recommended using a patent co-word map analysis (PCMA) in order to assess the tendencies of technological trends in the field of hydrogen energy. Furukawa et al. (2015) provided a method to analyze chronological changes in research themes as seen from proceedings articles and conference sessions in order to discover, identify, and analyze the evolutionary process of new technologies in the numerous rapidly expanding research domains. Joung and Kim (2017) applied a keyword-based model in contents-based patent analysis and suggest a technical keyword-based analysis of patents to track developing technologies.

Weismayer and Pezenka (2017) offered a longitudinal latent semantic analysis of keywords as an application to content analytics.

In recent years, lexical-based approaches mostly mixed with other approaches, especially indicator-based approaches to detect emerging technologies. Therefore, a method that only uses lexical-based approach is hard to find in previous studies.

### 2.2.2. Bibliometric-based approaches

We group the approaches that are mostly based on the relationship between papers or patents, and citation analysis between them. Daim et al. (2006) was one of the first works that used this approach by combining the use of bibliometrics and patent analysis with well-known technology forecasting methods including scenario planning, growth curves, and analogies for three emerging technological sectors.

Shibata et al. (2009) performed three citation network methods to detect a research front including co-citation, direct citation, and bibliographic coupling, and found out that direct citation performs the best to detect large and young clusters earlier. Shibata et al. (2008) successfully used topological measures for detecting branching innovation in the citation network of scientific publications. Shibata et al. (2011) detected emerging research fronts and future core papers using the topological clustering method and citation network analysis. Chen et al. (2011) deployed the logistic growth curve approach to detect emergence, growth, maturity, and saturation in the field of hydrogen energy using patents.

Iwami et al. (2014) identified emergent leading papers using time transition of centrality measures. Yoon and Kim (2012) used outlier patents as sources of emergent technologies and semantic patent analysis as sources of topics. Kim and Bae (2017) formed technology clusters and with the usage of patent indicators assessed whether or not a technology cluster is promising or not.

In order to suggest potential future research topics for technological observatories, Santa Soriano et al. (2018) examined citation patterns and co-occurrence keywords while evaluating their importance and level of maturity. Mejia and Kajikawa (2020) used a computational algorithm based on citation networks and thoroughly examined energy storage emerging topics by mining journal articles and patents.

### 2.2.3. Indicator-based approaches

Porter and Detampel (1995) was one of the first works that used the number of records that include a specific keyword in their abstract as an indicator of emerging technologies to detect in bibliometric databases. Watts and Porter (1997) introduced five indicators of Research and Development(R&D) profiles: the number of items in databases such as Science Citation Index as Fundamental research, the number of items in databases such as Engineering Index as Applied research, the number of items in databases such as U.S. Patents as Application, the number of items in databases such as Newspapers Abstracts Daily as Fundamental research, issues raised in the Business and Popular Press Abstracts as Societal Impacts. Bengisu (2003) used the slope of the regression line of the number of records in the specific field to the time as an indicator of emergence in fields. Watts and Porter (2003) defined some cluster quality measures including cohesion (as the cosine similarity measure), entropy, and F-measure as emergence indicators. Bettencourt et al. (2008) used an epidemic model to relate the increasing number of publications and new authors in an emerging field. Schiebel et al. (2010) used good indicators such as TF-IDF (term frequency-inverse document frequency) scores and Gini coefficient, as well as the minimum number of articles that contain the keyword to detect an emergent research issue. The former study can be considered as one of the first ones that detected emerging topics with multi-layer filtering approaches and indicators. Guo et al. (2011) deployed three indicators of the number and type of bursting terms, the number of new authors in a field, and the interdisciplinary paper references to identify emerging research areas. Järvenpää et al. (2011) used Technology Life Cycle indicators based on the databases that have been used in emerging technologies detection including the number of articles in the science datasets, the number of patents in the patent datasets, and the amount of news in newspaper datasets. Abercrombie et al. (2012) constructed a network of scholarly publications, citations, patents, news, and online mappings to discover the relations of the indicators of each, for any emerging technology. Jun (2012) evaluated the technology hype cycle of hybrid cars and used google search traffic or Google trend as an indicator of users' behavior. Jun et al. (2014) believed that search traffic (Google trend) can be a better measurement of new technology adoption than other indices such as patents, news, or articles for forecasting demands. De Rassenfosse et al. (2013) believed that the inventor's total number of priority patent applications, regardless of the patent office where they were submitted, can be an indicator to assess and detect emerging technologies with. Ho et al. (2014) used a fitted logistic curve on the cumulative number of publications in a field per year to assess the emergence and predict the life cycle of that technology.

H. Xu et al. (2021) used eight indicators to detect emergence, including average growth rates of paper numbers, journal numbers, funding numbers, authors numbers, weakly connected components, strongly connected components, plus publications cited by patents divided by the total number of publications on a topic, and patents cited by publications divided by the total number of publications on a topic.

#### 2.2.4. Machine learning approaches

Because of the successful usage of machine learning approaches in different fields, the implementation of machine learning approaches has increased in the field of emerging technology detection as well. S. Xu et al. (2019) used the Dynamic Influence Model (DIM) to detect topics and then by using Citation Influence Model (CIM), they calculated input indicators and predicted the next two years' values with Multi-Task Least-Squares Support Vector Machine (MTLS-SVM). Zhou et al. (2019) combined a semi-supervised text-clustering model (Labeled Dirichlet Multi Mixture) for topic segmentation and a sentence-level semantic description method (Various-aspects Sentence-level Description) information extraction method for topic description to identify emerging technologies using a semi-supervised topic clustering model. Zhou et al. (2020) built a supervised machine learning model to label them ET (Emerging Technology) or NET (Not Emerging Technology) and patent features as inputs with the usage of data augmentation with GAN (Generative Adversarial Networks) to build enough data to train the model. Altuntas et al. (2020) evaluated emerging candidates with the patent analysis using a semi-supervised clustering. Ma et al. (2021) proposed a hybrid approach to integrate topic modeling, semantic SAO analysis, machine learning, and expert judgment, identifying technological topics and potential development opportunities. Zhou et al. (2021) deployed 11 patent indicators to detect emerging technologies that are large-scale outlier patents using technological and social impact with the deep learning method. Jang et al. (2021) used expert opinions on future and emerging technologies identified through the LDA (Latent Dirichlet Allocation) model and fuzzy c-means probabilistic clustering by utilizing diversity and centrality indices. Choi et al. (2021) deployed three semi-supervised active learning algorithms with 32 input variables of patents and one binary target variable of being promising or not to identify emerging promising technologies.

#### 2.2.5. Hybrid approaches

In recent years more groups use the complex methods that mix different approaches we call hybrid approaches.

Ávila-Robinson and Miyazaki (2013) used bibliometric indicators to detect technological emergence. Q. Wang (2018) used bibliometrics and indicators of growth of the number of publications, the novelty of the topic, coherence of the cluster of a topic, and the number of citations as the scientific impact to detect an emerging research topic.

J. Garner et al. (2017) used a set of indicators to evaluate the emergence of the terms in terms of novelty, growth, community, and persistence, which was a combination of lexical and indicator methods; their method was called Emergence Score (EScore). Carley et al. (2017) used the same EScore method but evaluated the effects of scale and domain on the persistence of an emerging topic. Carley et al. (2018) elaborated on the EScore method more thoroughly and implemented it on a dye-sensitized solar cells (DSSCs) dataset and found the emergent terms, authors, and affiliations in this field. Porter et al. (2019) implemented the EScore method and revised it to identify the emerging terms and key players, as well as high-priority research papers and patents. Ranaei et al. (2020) evaluated and compared this method to other methods to see the strengths and weaknesses of the EScore method.

### 2.3. Keyword extraction

The method of automatically extracting relevant keywords from a document is known as automatic keyword extraction (AKE) (Nasar et al. 2019). The main purpose of AKE is to summarize the main ideas of a document in some keywords or key phrases. This is how it can be related to the idea of emergent topic detection; in which we are going to extract the terms related to a document that can have the potential of emergence based on some characteristics or indicators.

#### 2.3.1. Term clumping

Term clumping is a semi-automated process of cleaning, consolidating, and clustering the terms (Zhang et al. 2014); It has been used by some papers in the process of emergence detection (Carley et al. 2017, 2018; Garner et al. 2017; Huang et al. 2021; Liu and Porter 2020; Porter et al. 2019; Ranaei et al. 2020). Most of the previous studies saw the keyword extraction step as part of the term clumping, and they did not pay enough attention to the keyword extraction step or its different methods. However, we believe that the keyword extraction method can be particularly important and can play a vital role in the process of emergence detection.

#### 2.3.2. BERT

Bidirectional Encoder Representations from Transformers or BERT is one of the recent Natural Language Processing (NLP) models that has been proposed (Devlin et al. 2019). It is a representation model that can be used for the keyword extraction step. Ebadi et al. (2022) was the first study that used the BERT method for extracting keywords in the emergence detection process. Followed by that approach, we utilize the BERT method in our keyword extraction step of the emergence detection process.

### 2.4. Indicators of Emerging Technologies

#### 2.4.1. Growth

Cozzens et al. (2010) included recent fast growth as one of the main attributes of emerging technology. Small et al. (2014) believed that despite all the different views on emerging technologies definitions and attributes, there is a consensus that growth and novelty are two properties related to emergence. A term to be considered an emerging technology should have a relatively fast growth (Rotolo et al. 2015). Growth can be related to actors such as organizations, authors, or users and outputs such as publications, patents, and products (Rotolo et al. 2015). Porter and Detampel (1995) used growth in the number of keywords in publication titles and abstracts as an indicator of emergence. After that, many studies used the growth in the number of records including a topic as an emergence indicator of that topic (Carley et al. 2017, 2018; Garner et al. 2017; Porter et al. 2019; Q. Wang 2018). Some used the growth of the number of records of a topic as an indicator of growth (Bengisu 2003; Small et al. 2014; Q. Wang 2018; H. Xu et al. 2021). One of the differences is how we count the keywords; binary counting counts a term based on the presence or absence of a term in a document, and full counting counts all the occurrences of a term in a document. Most of the previous studies use binary counting but full counting can be meaningful as well.

**Short-term growth:** Guo et al. (2011) believed that a burst of keyword usage is needed for a specific term to be emergent, which means it should have a short-term fast growth. Cozzens et al. (2010) emphasized the recent fast growth as well, which means it has to have high short-term growth.

**Long-term growth:** Previous studies used a kind of long-term growth in their emergence scores (e.g., Porter et al. 2019). It seems reasonable because, in the long term, an emergent term would have a positive growth rate. Long-term growth can be related to both persistence and novelty. It is related to persistence because an emerging term should be active and probably with growing potential for several years. It is related to novelty because if the topic is new, it should not have an active period before that, which can result in a positive long-term growth rate.

### 2.4.2. Novelty

Cozzens et al. (2010) listed newness (novelty) as one of the key characteristics of emerging technologies. Small et al. (2014) believed that despite all the different views on emerging technologies definitions and attributes, there is a consensus that growth and novelty are two properties related to emergence. A term to be considered as an emerging technology should have radical novelty (Rotolo et al. 2015). Q. Wang (2018) also believed that the number of publications at the beginning of a period should be low, and it should have radical novelty at the first stage of its emergence period.

**Short-term novelty:** The concept of novelty or newness requires that the records that work in this field be low. Therefore, the number of records should not exceed a minimum number. However, it should be noted that setting an absolute value is risky because it might be affected by many factors such as the size of the dataset and the field of a topic. So, it should be moderated by other factors.

**Long-term novelty:** One way to evaluate novelty is by measuring the number of records in the first years of the emergence and some years later. Despite the absolute value, this relative value seems more reliable because every emergent topic during the first years of its emergence has a much smaller number of records in comparison to the upcoming years. The problem is that this type of novelty measurement can be similar to the growth measurement as if we do not calculate the novelty, we calculate the growht.

### 2.4.3. Community

In some studies, the attribute of community was selected instead of coherence to show that the connections between the other groups or clusters are not loose (e.g., Ranaei et al. 2020). The coherence attribute is more appropriate for the clustering-based methods, therefore, substituting it with another attribute is logical for other methods. However, by the definition and the way of calculation based on J. Garner et al. (2017), the community attribute is not effective. Their criterion for the community is that the term is "required to have more than one author that doesn't share the same record set" (Garner et al. 2017). Because the other constraints are much more powerful than this constraint, it will hardly be activated. Therefore, it is not actually a part of the emergence detection process. Instead of the attributes of community or coherence in the previous studies, we believe an indicator is required that can be calculated by the number of active players such as authors, organizations, and countries. So, we choose collaboration as our indicator to show the coherence of a topic to be connected to various groups of active players.

Although the first steps of creating a research topic or science collaboration might be expected to be less, because the trust can be low (Mund and Neuhäusler 2015), for growing a field or topic people need to help each other to raise a field. Without collaboration, even the best ideas cannot convert into ground-breaking phenomena. Collaboration as an indicator was first used by Abercrombie et al. (2012) but it was only about the number of patents per year by a country or international collaboration; also, it was not used as part of a method for the detection of an emerging technology or topic. Ávila-Robinson and Miyazaki (2013) defined two indicators of the collaboration at the country level and the collaboration at the author level to measure the level of collaboration in the knowledge generation. They also found out that the emerging science topics have higher levels of collaboration. So, collaboration can play an important role in the process of emerging topic detection. If a topic is emergent, some collaborations should be formed around that topic. As it grows, the collaboration around that topic should grow as well. Some previous studies included growth in collaborative actors of emerging technologies, such as growth in the number of authors and organizations, in the growth dimension of emergence (Rotolo et al. 2015; Xu et al. 2021).

#### 2.4.4. Impact

Rotolo et al. (2015) defined prominent impact as one of the main attributes of emerging technologies. Emerging technologies can "yield benefits for a wide range of sectors" (Martin 1995), "exert much enhanced economic influence" (Porter et al. 2002), or change "the basis of competition" (Hung and Chu 2006). Emerging technologies have the potential to influence significantly by permeating several socioeconomic system levels, including organizations, institutions, knowledge creation processes, and technology regimes (Xu et al. 2021). Previous studies had different views on the impacts of emerging technologies. Although other impacts like societal or economic impacts are important we focus on the scientific and technological impacts of emerging technologies as the main attributes of emerging technologies.

**Scientific impact:** Q. Wang (2018) considered scientific impact as one of the main attributes of emerging research topics. Its measurement for scientific impact was the number of citations for a targeted research topic in a specific period. This scientific impact can be a good sign of early emergence because at first, only high-quality and pioneer papers about a topic will be published. Therefore, the quantity is not high, but the quality is high, and the average citation number would expect to be high. Then, after the emergence period, everyone tries to enter the topic and the average citation is going to come down.

**Technological impact:** Day and Schoemaker (2000) believed that emerging technologies can create a new industry or change the previous ones. We believe that one of the main impacts of emerging technologies is the technological impact. There are various studies in the literature that use patent datasets to detect emerging technologies, but few research works employed patent databases and patent counts as attributes of technological impact. (Watts and Porter 1997). On the other hand, Abercrombie et al. (2012) constructed a network of scholarly publications, citations, patents, news, and online mappings to discover the relations of the indicators of each, for any emerging technology. Jun (2012) evaluated the technology hype cycle of hybrid cars, and Jun et al. (2014) used google search traffic or Google trend as an indicator of users 'behavior.

### 2.4.5. Persistence

Persistence is also another main attribute of emergence, and it is called the ability to stay in power (Carley et al. 2017). Carley et al. (2018) detected persistence by passing two constraints of having at least 7 records in 10 years and appearing in at least 3 time periods.

## 2.5. Emergence Score

Measurement of emergence has been attempted using a variety of methods. The scope of this effort might range from assessing macro-level factors – such as national propensities (Porter et al. 2002)- to micro-level scientific problems (Boyack et al. 2014), and from monthly assessments to decades-long time periods (Lacasa et al. 2003). Small et al. (2014) described an innovative method for leveraging literature data to identify new science and technology (S&T) themes. They used direct citation and co-citation to identify thematic clusters and follow temporal patterns using the Scopus database. J. Garner et al. (2017) looked for emerging themes within certain study domains (a micro approach). They presented a method based on Rotolo et al. (2015) for detecting emerging technologies or topics with a step-by-step quantitative method but in a micro approach. They performed emergent technology detection on specific domains such as dye-sensitized drug delivery, nano-enabled drug delivery, big data, and non-linear programming datasets successfully. Some studies also were performed with the same datasets but with the intention of exploring more about the same results (Carley et al. 2017, 2018; Porter et al. 2019).

J. Garner et al. (2017) used a set of indicators to evaluate the emergence of the terms in terms of novelty, growth, community, and persistence with an easy to implement and interpret but effective method, which we also selected as our basic pipeline. The emergence score revealed by Z. Wang et al. (2019) and contained three elements of active trend, recent trend, and slope, all of which were mostly related to the attribute of growth. Porter et al. (2019) deployed the same EScore method for detecting emerging technologies and players and applied thresholds to meet novelty, persistence, and community criteria; however, they believed that EScores only reflect growth patterns. The EScore cannot measure the degree of metrics such as novelty, persistence, and community (Porter et al. 2019). Their method was promising, but they used some limited metrics to have a threshold and did not use that metric to evaluate the final emergence score which makes it difficult to implement and decipher. In other words, that study has limited results to make implications. Despite the drawbacks, noted method can be implemented by both paper and patent datasets (Z. Wang et al. 2019).

Q. Wang (2018) also used a set of attributes including novelty, growth, coherence, and scientific impact to detect emerging research topics. The measure can provide a score for each attribute of emergence but cannot assign an emergence score for each emergent research topic. It also uses some thresholds to detect the emergence, and an expert should set those values. Therefore, the detection process and how many emergent topics will be detected is sensitive to the threshold values. This method has the disadvantage of not being applicable for detecting emergent technologies through patents because it uses a method for measuring scientific impact for papers and is designed to detect emergent research topics.

As it can be deducted, providing a comprehensive approach that can use different emergence attributes and uses them in the process of detecting emergence and also measuring emergence is necessary; an approach that can be used to detect emerging topics prevalent both in academia and industry.

## 3. Data

Data were extracted from the WOS, using the following search query: ("artificial intelligence*" OR "neural network*" OR "machine learning*") in their title or abstract and for the period from 2010 to 2019. We extracted the most relevant data from them, resulting in 4992 papers. We used the same query and the time span of 2018 and 2019 to extract 5000 patents from the Derwent Innovation database as well. The type of records in the paper dataset can be seen in Figure 2. There is an increasing trend in the number of papers representing the growing field of AI in the scientific community.

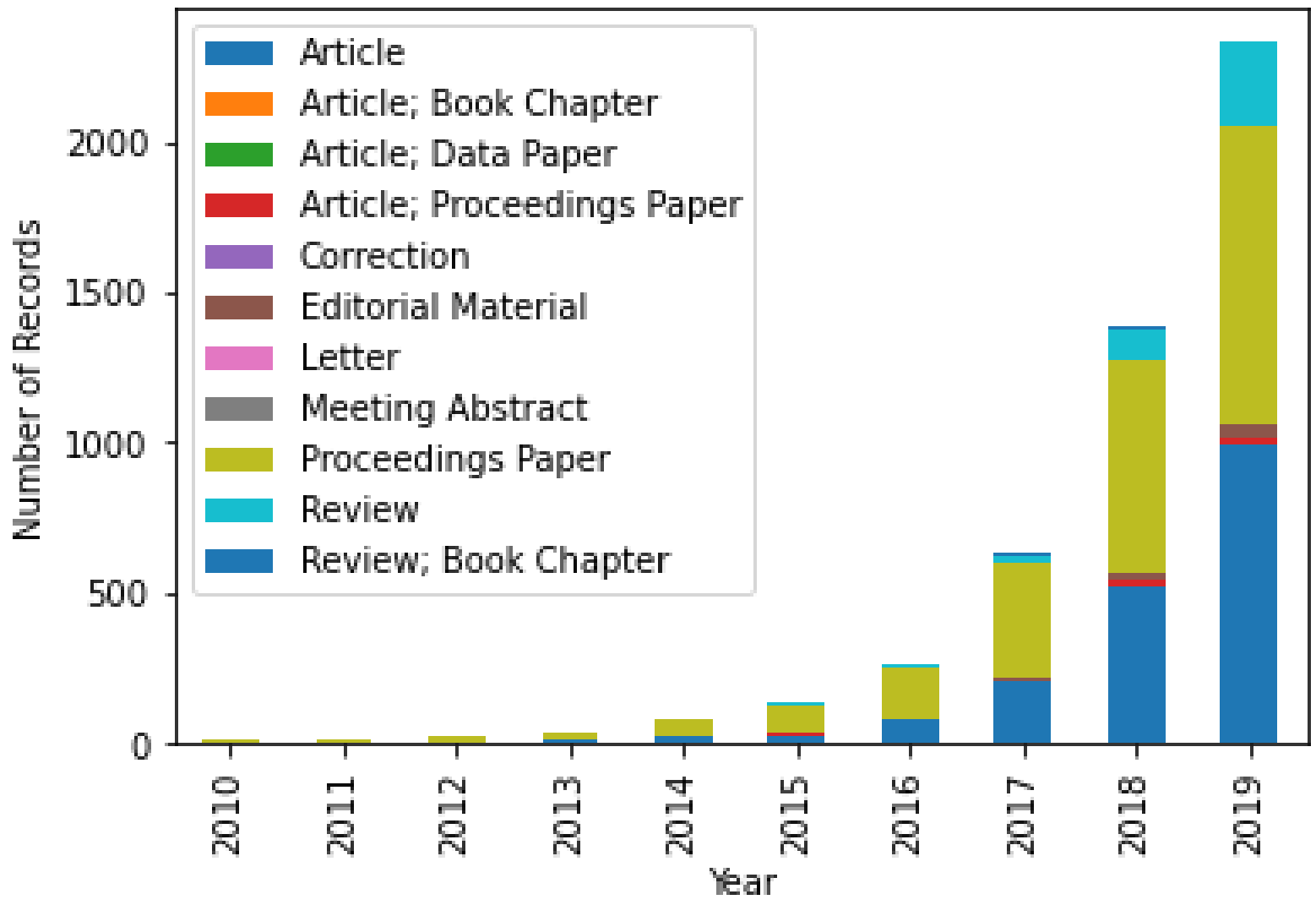

**Figure 2.** The Annual distribution of the publications in the WOS dataset

## 4. Methodology

The whole conceptual flow of the analysis is depicted in Figure 3. As seen in this figure, we divided the whole work into five steps: data collection, text processing, keyword extraction, emergence metrics evaluation, and emergence score calculation. The data was first collected from two databases and then preprocessed by dropping not-a-numbers (NaNs) and removing null and duplicate items. Then, text processing steps including lower case conversion, punctuations removal, tokenization, lemmatization, numbers removal, and stop words removal were applied. The BERT keyword extraction method was then deployed to extract the potential noun phrases. After that, several emergence metrics were calculated based on the attributes of growth, novelty, collaboration, and impact. Finally, based on the emergence metrics and attributes, the emergence scores were calculated.

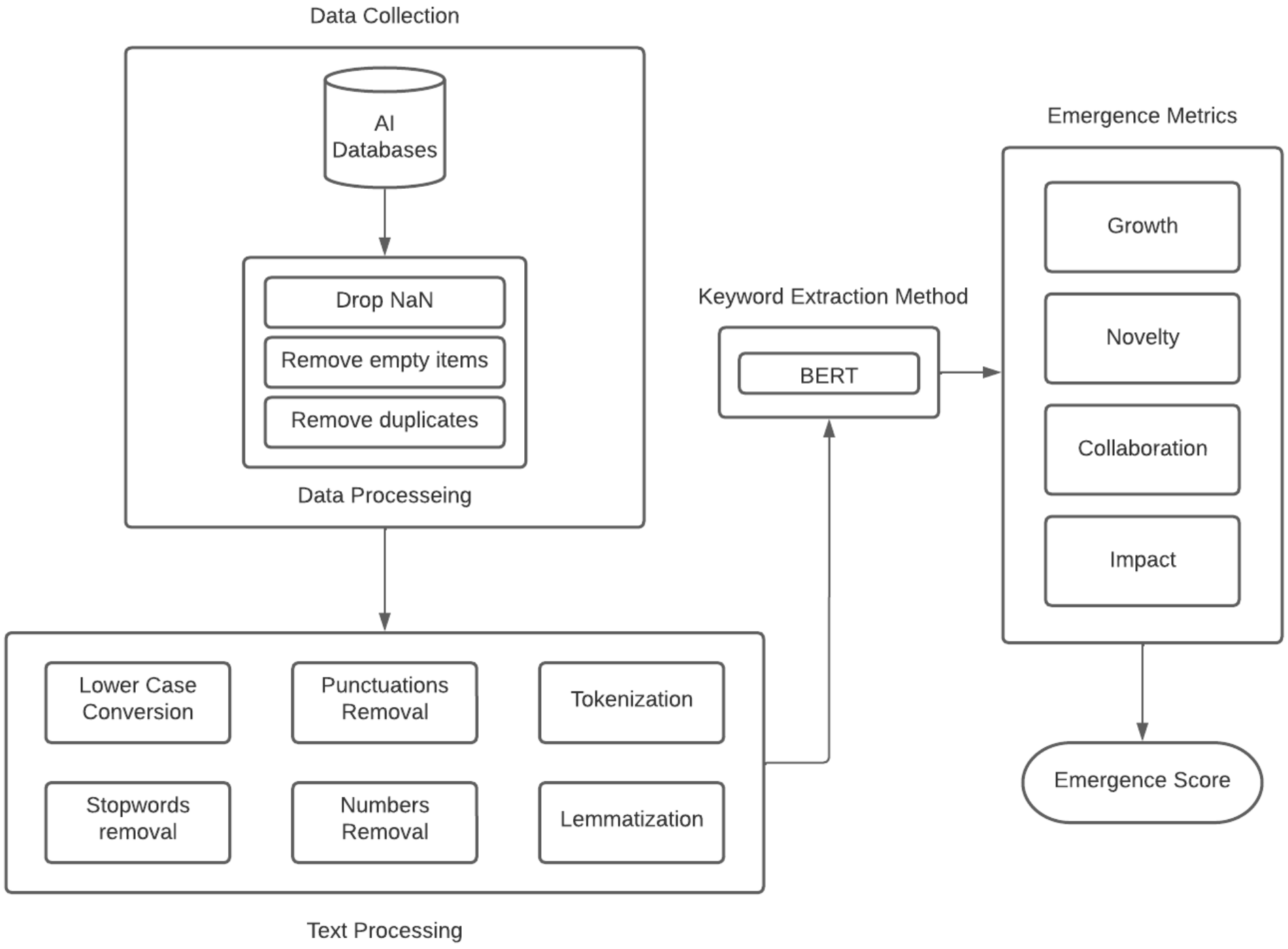

**Figure 3.** Workflow of the proposed method

For detecting emerging technologies, we rely on emergence attributes. The attributes and their evaluation should come from the definition of emerging technology and its nature. We assume the following process is logical for an emerging topic: people start to study and work on a topic in science. At first, there are not a lot of researchers working on this topic. Little by little, more researchers read about this new topic and attract to it. Therefore, it has novelty at first, which means it is new and not many researchers work on that, and it has grown in the middle, which means more are attracted to this new topic, which can lead to more and more collaboration on this topic. However, it is not the whole story. An emerging topic is expected to gather scientific and technological impact that can be indicated through the paper that cites this topic and the patent that appears on this topic.

We modify the previous attributes because we believe that the persistence attribute can be depicted with the impact attribute. An emerging topic has some impacts. These impacts are the reason that an emerging topic can have persistence and be continued. On the other hand, other attributes such as growth can have the persistence attribute in themselves very well, because if an emerging topic has shown a great deal of growth, it is probably having enough amount of persistence. Therefore, new and previous attributes of emergence can play the role of persistence pretty well.

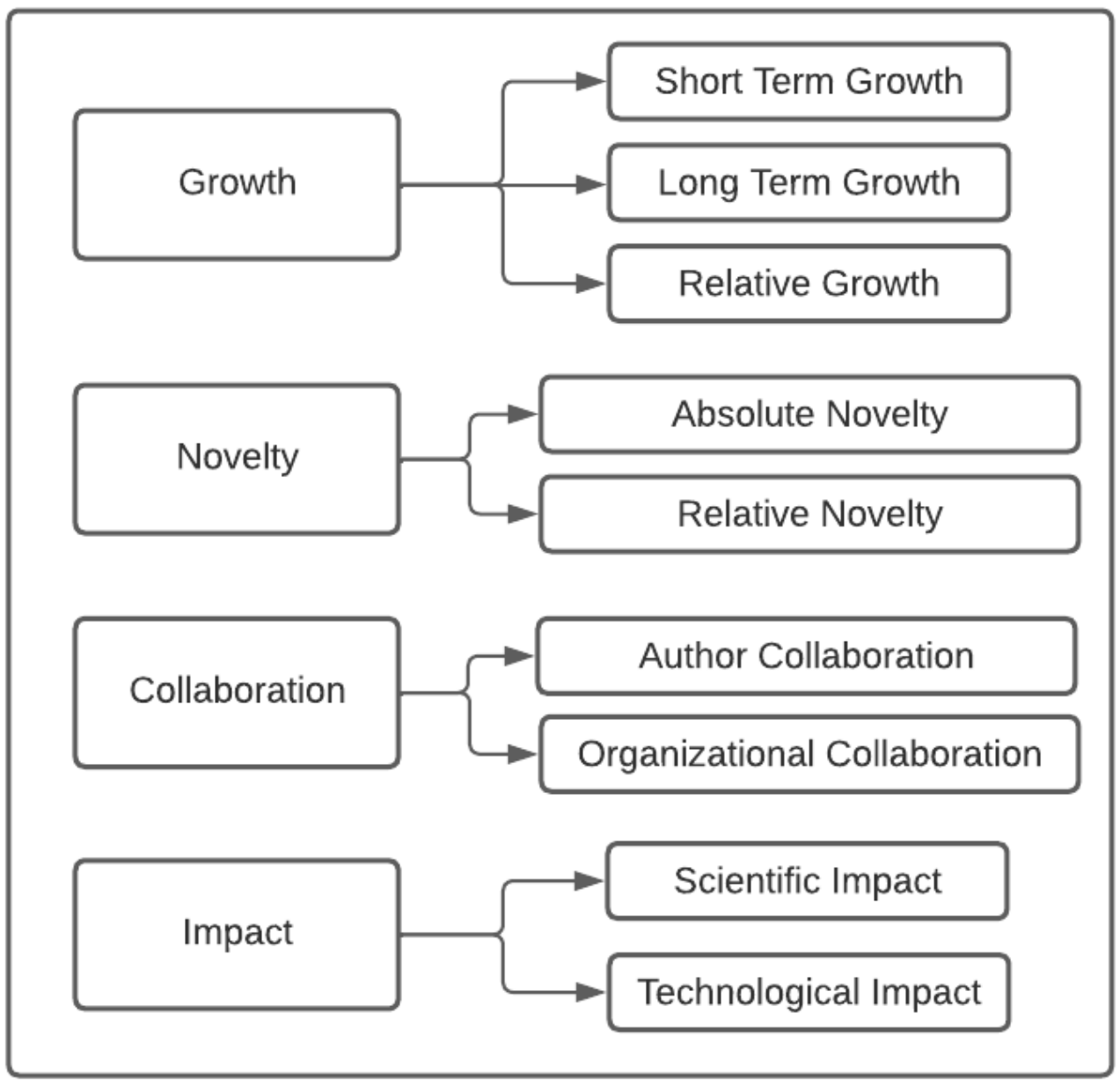

**Figure 4.** Metrics deployed in our method

All the attributes and the metrics that we used are depicted in Figure 4. We will elaborate on each attribute and metric briefly in the following.

## 4.1. Growth

### 4.1.1. Long-term growth

We measure long-term growth as the slope of the number of papers in the long-term of the study.

As noted earlier, long-term growth is one of the most important metrics to evaluate the emergence, because an emergent topic is expected to have a very high long-term growth.

We defined long_term_growth as the natural logarithm of the slope of change in the number of papers active in the field in the last 3 years to the first 3 years of the study.

We also defined long_term_growth2 as the natural logarithm of the slope of change in the number of papers active in the field in the last 5 years to the first 5 years of the study.

### 4.1.2. Short-term growth

We measure short-term growth as the slope of the number of papers in the short term of the study.

Short-term growth is also an important factor to detect emergence. It can distinguish between an emergent topic and a hot-topic because many hot-topics still have long-term growth but not a very high short-term growth. An emergent term is expected to have both long-term growth and short-term growth.

We defined short_term_growth as the natural logarithm of the slope of change in the number of papers active in the field in the most recent 2 years and the prior 2 years of the study.

We also defined short_term_growth2 as the natural logarithm of the slope of change in the number of papers active in the field from the mid-year of the active period to the most recent year.

#### 4.1.3. Relative growth

We measure relative growth as the growth rate of the percentage of the papers active in a field to all papers in years between 5 and 10.

It is an important metric that has been forgotten to take into account. It is expected to see an increasing number of papers in a field or domain. Therefore, if we see a long-term and short-term trend for a topic, it does not necessarily mean its growth can be considered rapid growth, which is expected for an emerging topic. Many topics still get attention and have positive growth rates, but they are mostly hot-topics not emerging topics. We believe that an emerging topic should have a relative growth rate higher than the other hot-topics and trends. For instance, “machine learning” is expected to have a high growth rate in the last decade, but the percentage of papers active in this field is not expected to grow dramatically in the last decade. However, an emerging topic should have a high rate of growth, not only in the number of papers but also in the percentage of papers in an emerging topic that should grow drastically.

### 4.2. Novelty

#### 4.2.1. Absolute novelty

The concept of novelty or newness requires that the records that work in this field should be low. Therefore, the number of records should not exceed a minimum number. However, it should be noted that setting an absolute value is risky because it can be affected by many factors such as the size of the dataset and the field of a topic. It can be moderated by other factors. One way to evaluate novelty is the measurement of the number of records in the first years of the emergence. We set the first two years of the study to calculate the absolute novelty of a topic.

#### 4.2.2. Relative novelty

Despite the absolute value, this relative value seems more reliable because every emergent topic in the first years of its emergence has a much smaller number of records in comparison to the upcoming years. Therefore, we see the ratio of active papers on a topic to all papers in a period. We set the first three years of the study to measure the relative novelty of a topic.

Relative novelty:

novelty_score = $5 * e^{-10x}$

x = The percentage of the number of papers active in a topic to all papers in the first three years of the study

Absolute novelty:

novelty_score_2 = $5 * e^{-x}$

x = The total number of papers active in a topic in the first two years of the study

### 4.3. Collaboration

#### 4.3.1. Author's collaboration

- The number of authors active in a field.
- The number of author groups active in a field.

An emerging topic should gather groups together. One of the most important actors is the authors. If one topic is only attracted a few authors or author groups, it does not have the coherence that is expected for an emergent topic. On the other hand, the more authors and author groups gather around a topic, the more emergence potential it has.

We defined collab_author1_growth as the natural logarithm of the slope of change in the number of authors active in the field in the last 3 years to this previous period.

We also defined collab_author2_growth as the natural logarithm of the slope of change in the number of authors active in the field in the last 6 years to the first 4 years of the study.

#### 4.3.2. Organization's collaboration

We measure organization's collaboration as the number of organizations active in a field.

Another important actor is an organization. If a topic can attract different organizations, it seems to have more potential to have an impact on the economy and society. Therefore, the organization's collaboration has the ability to reflect the emergence very well.

We defined collab_organ_growth as the natural logarithm of the slope of change in the number of organizations active in the field in the last 4 years to the first 6 years of the study.

### 4.4. Impact

#### 4.4.1. Scientific impact

We measure scientific impact as the average number of citations per year of a topic.

An emerging topic is expected to have a higher scientific impact. Because at first, few but high-quality papers introduce a new topic, and then these topics are read and cited by other researchers. Therefore, it is expected that the emergent topics can have more average citations than the non-emergent topics.

We defined technological_impact as the natural logarithm of the average number of citations per year of a topic in the whole period of the study.

#### 4.4.2. Technological impact

We measure technological impact as the number of patents including the topic in their title or abstract in the last two years.

The technological impact is one of the attributes that is directly related to the definition of emergent technologies. An emergent topic is expected to have the potential to be used in technologies and one of the best indicators that a topic is being used in technologies is the usage of that topic in patents. Because of the essence of technological impact, there is a delay between introducing a topic in academia and using it in technology. Therefore, we evaluate the usage of a topic in the last two years of our study to detect the technological impact of a topic.

We defined technological_impact as the natural logarithm of the number of patents including the topic in their title or abstract in the last two years.

### 4.5. Emergence_Score

We finally calculate the Emergence Score with the indicators that we defined previously.

One of the major challenges that inhibit the researchers to have a numeric score for emergence is that emergence attributes are different, and emergence metrics are not totally comparable. In this study, we tried to normalize the emergence metrics in a way that they can finally produce a comparable score of emergence. It is done through different computational operations from using natural logarithm or exponential growth to using absolute values and sometimes average of some attributes. Finally, we believe that the proposed attributes and metrics with the special operations that we used can provide us with a reliable and comparable emergence score that is presented as follows.

Emergence_Score = collab_author_growth1 + collab_author_growth2 + collab_organization_growth1 + relative_growth + technological_impact + scientific_impact + novelty + novelty2 + long_term_growth + long_term_growth2 + short_term_growth + short_term_growth2

## 5. Results

Table 1 shows the AI emergent terms with the highest emergence scores. These terms were approved by a domain expert.

**Table 1.** Emergent terms with highest emergence scores

| AI Emergent term | Emergence Score |
|---|---|
| convolutional neural network | 63.05 |
| deep reinforcement learning | 55.40 |
| machine learning models | 53.18 |
| medical imaging | 51.90 |
| big data | 51.56 |
| IoT | 51.44 |
| random forest | 51.22 |
| image analysis | 51.21 |
| deep learning models | 51.11 |
| clinical practice | 50.55 |
| lstm | 50.16 |
| deep learning technology | 49.02 |
| recurrent neural network | 48.62 |
| health care | 47.96 |

Table 2 shows the AI emergent terms with the highest EScores; a method proposed by J. Garner et al. (2017). There is almost a significant difference between the emergent terms among the two methods. The AI expert approved that the output terms of our proposed approach are more accurate to be considered emergent terms from 2010 to 2019. The convolutional neural network was one of the main emergent technologies in AI in this decade, while the other technologies that were the output of the previous EScore method were almost continued emergent technologies of previous years not emergent technologies of those years.

**Table 2.** Emergent terms with highest EScores in the EScore method proposed by Carley et al. (2018)

| AI Emergent term | EScore |
|---|---|
| machine learning | 226.88 |
| deep learning | 205.01 |
| neural networks | 83.28 |
| artificial intelligence technology | 71.01 |
| artificial neural networks | 67.80 |
| artificial intelligence algorithms | 66.59 |
| machine learning algorithms | 59.67 |
| artificial intelligence techniques | 58.22 |
| decision making | 55.58 |
| machine learning techniques | 53.77 |
| artificial intelligence methods | 51.54 |
| deep learning techniques | 49.54 |
| artificial intelligence systems | 49.11 |
| computer vision | 48.06 |

Table 3 illustrates the output of another method of emergence detection, the maximum co-occurrence between terms. This method also provides results that are useful but not emergent. The output of this approach can mostly relate to the hot-topics and trend topics of the period of study. Figure 5 also provides us with a term co-occurrence map and the relationship between different term clusters.

**Table 3.** Top terms with maximum co- occurrences based on VOSviewer

| Keyword | Occurrences | Total link strength |
|---|---|---|
| deep learning | 2537 | 9128 |
| machine learning | 2021 | 7945 |
| artificial intelligence | 1563 | 5605 |
| classification | 542 | 3199 |
| neural-networks | 330 | 1905 |
| prediction | 258 | 1612 |
| big data | 233 | 1280 |
| algorithm | 199 | 1104 |
| model | 192 | 1090 |
| system | 138 | 819 |
| diagnosis | 120 | 815 |
| convolutional neural network | 175 | 755 |
| segmentation | 100 | 725 |
| support vector machine | 129 | 622 |

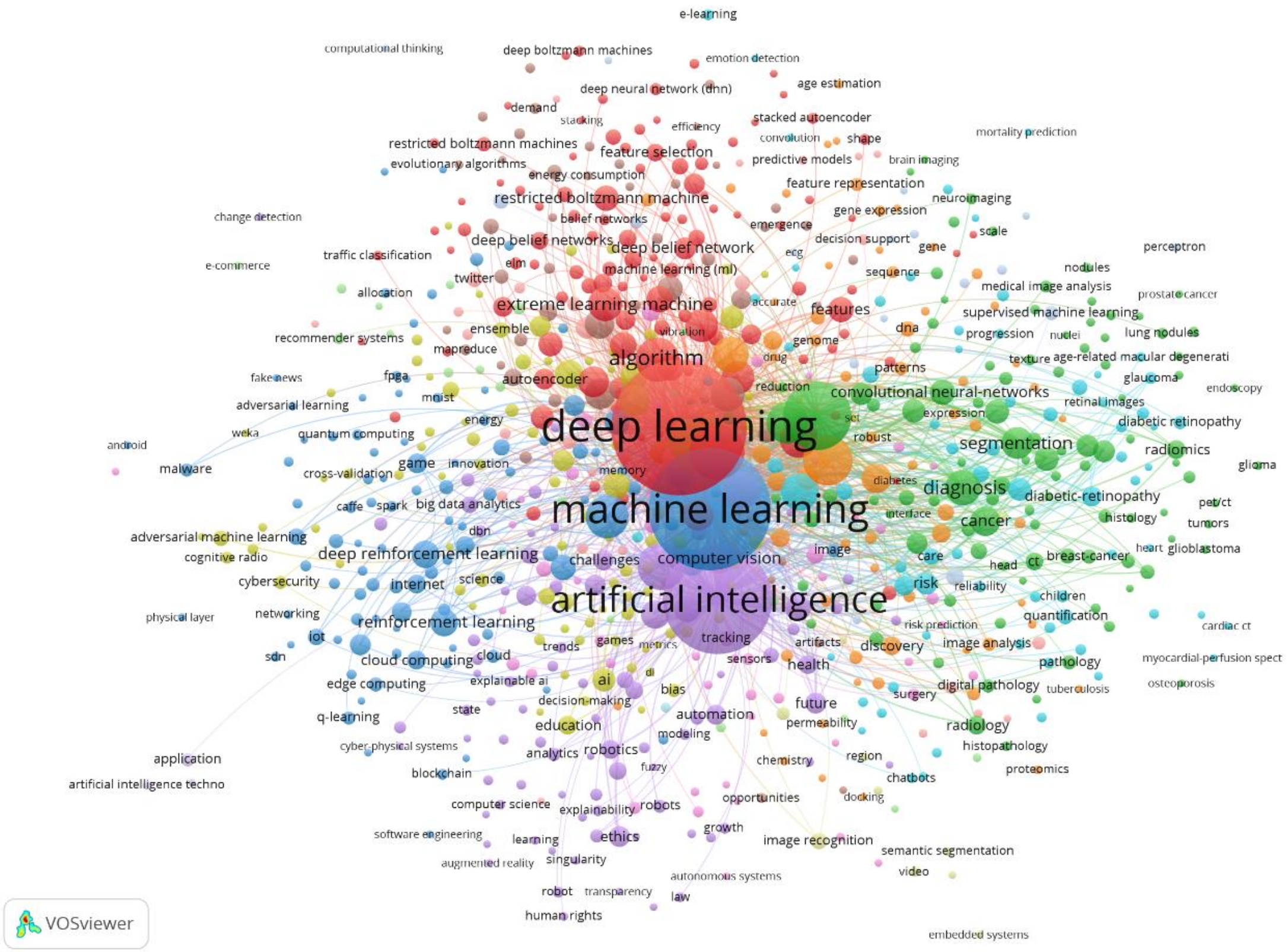

**Figure 5.** Term co-occurrence map in AI from 2011 to 2019

We also added the results of a simple TF-IDF method. The TF-IDF scores are useful as well, but cannot detect the emergence and early detection of emergence that is useful for the process of identification of emergent technologies.

**Table 4.** Top terms with maximum TF-IDF scores

| # | Emergent Term | Score |
|---|---|---|
| **1** | deep learning | 215.69 |
| **2** | artificial intelligence | 174.89 |
| **3** | machine learning | 174.59 |
| **4** | classification | 95.68 |
| **5** | neural network | 73.59 |
| **6** | big data | 43.16 |
| **7** | convolutional neural network | 42.76 |
| **8** | support vector | 33.73 |
| **9** | reinforcement learning | 32.11 |
| **10** | machine learning algorithms | 29.75 |

In Figure 6 and Table 5, the correlation between our proposed metrics can be seen. It can provide researchers with useful information. For example, between the metrics and attributes, the novelty attribute has the least correlation with others. It is expected because novelty and other attributes are contradictory in their core definition. Or growth in organizational collaboration is highly correlated with the growth in author collaboration because most people are members of organizations and if people build a collaboration, they can build organizational collaboration as well. Another interesting attribute is the technological impact that has also one of the lowest correlations among the other attributes. Just like the novelty, it can also be different in the nature of its definition and measurement; it can get the effects of emergence with some time lag, and if a topic is growing it does not necessarily mean it will lead to the technological impact.

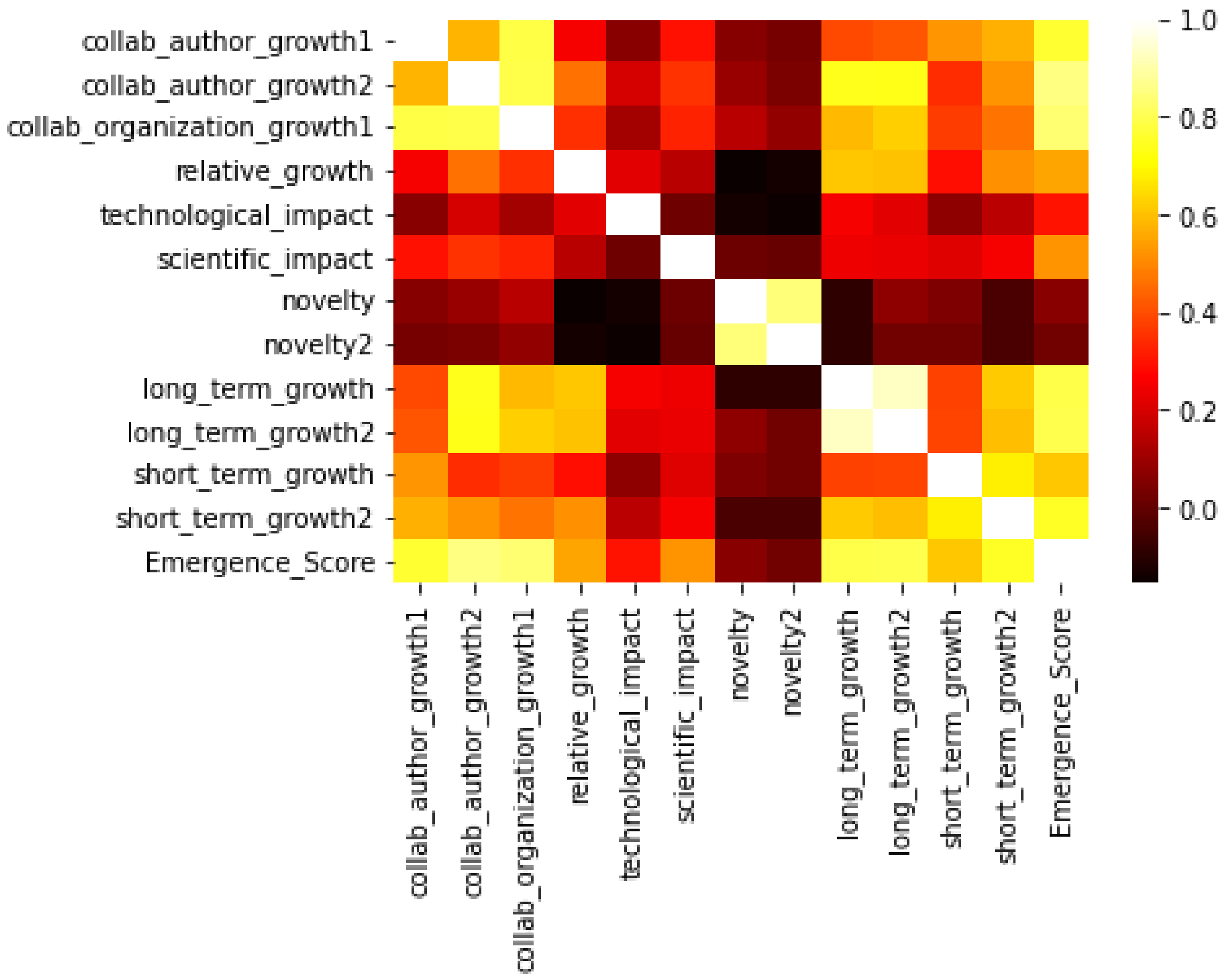

**Figure 6**. Correlation heatmap between metrics

**Table 5**. Correlation analysis between metrics

| Metrics | collab_author_growth1 | collab_author_growth2 | collab_organization_growth1 | relative_growth | technological_impact | scientific_impact | novelty | novelty2 | long_term_growth | long_term_growth2 | short_term_growth | short_term_growth2 | Emergence_Score |
|---|---|---|---|---|---|---|---|---|---|---|---|---|---|
| collab_author_growth1 | 1 | 0.58 | 0.78 | 0.25 | 0.06 | 0.29 | 0.06 | 0.03 | 0.39 | 0.41 | 0.52 | 0.57 | 0.76 |
| collab_author_growth2 | 0.58 | 1 | 0.79 | 0.46 | 0.2 | 0.35 | 0.09 | 0.04 | 0.74 | 0.73 | 0.35 | 0.52 | 0.85 |
| collab_organization_growth | 0.78 | 0.79 | 1 | 0.35 | 0.11 | 0.33 | 0.15 | 0.08 | 0.59 | 0.62 | 0.37 | 0.47 | 0.84 |
| relative_growth | 0.25 | 0.46 | 0.35 | 1 | 0.22 | 0.15 | -0.15 | -0.13 | 0.61 | 0.6 | 0.29 | 0.52 | 0.55 |
| technological_impact | 0.06 | 0.2 | 0.11 | 0.22 | 1 | 0.02 | -0.13 | -0.15 | 0.25 | 0.22 | 0.07 | 0.15 | 0.3 |
| scientific_impact | 0.29 | 0.35 | 0.33 | 0.15 | 0.02 | 1 | 0.02 | 0.01 | 0.24 | 0.23 | 0.21 | 0.26 | 0.52 |
| novelty | 0.06 | 0.09 | 0.15 | -0.15 | -0.13 | 0.02 | 1 | 0.84 | -0.09 | 0.07 | 0.05 | -0.04 | 0.07 |
| novelty2 | 0.03 | 0.04 | 0.08 | -0.13 | -0.15 | 0.01 | 0.84 | 1 | -0.09 | 0.02 | 0.02 | -0.04 | 0.02 |
| long_term_growth | 0.39 | 0.74 | 0.59 | 0.61 | 0.25 | 0.24 | -0.09 | -0.09 | 1 | 0.93 | 0.38 | 0.61 | 0.79 |
| long_term_growth2 | 0.41 | 0.73 | 0.62 | 0.6 | 0.22 | 0.23 | 0.07 | 0.02 | 0.93 | 1 | 0.38 | 0.6 | 0.8 |
| short_term_growth | 0.52 | 0.35 | 0.37 | 0.29 | 0.07 | 0.21 | 0.05 | 0.02 | 0.38 | 0.38 | 1 | 0.68 | 0.61 |
| short_term_growth2 | 0.57 | 0.52 | 0.47 | 0.52 | 0.15 | 0.26 | -0.04 | -0.04 | 0.61 | 0.6 | 0.68 | 1 | 0.75 |
| Emergence_Score | 0.76 | 0.85 | 0.84 | 0.55 | 0.3 | 0.52 | 0.07 | 0.02 | 0.79 | 0.8 | 0.61 | 0.75 | 1 |

We believe that previous methods of EScore mostly detect hot topics instead of emergent topics. Their methods of emergence score evaluation emphasized mostly on the growth. They cannot distinguish between a topic that has passed the emergence level and is growing naturally, which is a hot topic, and a topic that is in the emergence level and is growing at a bursting manner, which is an emerging topic. Adding the novelty measurement and some growth measurement, as well as the scientific impact, can make up for the previous drawbacks of EScore methods for detecting hot topics instead of emerging topics. Additionally, adding the technological impact to the emergence score measurement can make it more practical and converge the emergence score to its nature definition of emerging technologies which can have a profound impact on the economy and society.

We believe that for emerging technologies detection we mainly need to evaluate the novelty and growth parameters. Therefore, absolute values of papers on a topic in the first years should be low, and the second and third derivates of these values in the following years should be positive and high. However, the hot-topics can have a different trend. They do not need to have a low absolute number of papers on their topic or high second and third derivates of these values. If the number of papers on a topic is still high and ascending, it is enough to know that the topic is still hot and popular.

Another advantage of our method is that it does not base only on frequencies and occurrences of terms to determine the emergence degree. In previous methods in the literature, if a term frequency is increasing, it will have a positive emergence score, and the more its frequency and occurring, the more its emergence score (e.g., Porter et al. 2019). In that case, distinguishing between n-gram terms with shared words is difficult, and shared words with shorter lengths have higher chances to achieve emergence scores. For instance, if the trend is positive in the field of deep learning, the occurrences and probably slope of the term occurrences for the term “neural network” is higher than the term “convolutional neural network”, therefore, it will have a higher emergence score. However, in our proposed method, the occurrences and the slope of term occurrences are just some metrics, not the whole story. Adding other metrics, such as novelty, relative growth, scientific and technological impact had positive effects to overcome the problem of just paying attention to term frequency and its increase.

## 6. Conclusion

In this study, we tried to capture emerging topics in the field of AI for the period years from 2010 to 2019. In our proposed method, we considered two attributes of collaboration and technological impact plus some metrics such as relative growth. We tried to detect emerging topics with the usage of the attributes of novelty, growth, collaboration, and impact. To achieve this goal, we added some metrics for each attribute to have numerical values for these attributes. By the usage of considered attributes and metrics, we are able to have the final emergence score that is based on the attributes of emergence and can translate directly to each attribute. Because we have used the numerical values of each attribute, the final emergence score is more accurate than the previous methods that only use some predefined thresholds for each attribute and do not involve the emergence attribute in the process of emergence measurement. Especially, adding the novelty variable to the process significantly helped the method to early identify the emerging topics, not just reflecting some hot topics and trend topics in this period.

Also, having the patent information in our proposed method added a new layer of information that was missing elements in previous studies. Most previous studies only used patent or paper information, not both. However, by involving the patent information as a new attribute of technological impact we can detect information from academia and science but have the influence of technology and patents both in one method.

The proposed collaboration attribute can also detect the connection between researchers and organizations working on a topic. The previous methods in the literature used mostly clustering algorithms to evaluate the coherence between clusters as an attribute of emergence. With this proposed attribute, we no longer need to use those clustering methods. This attribute also helped us to better detect emerging topics with the degree of growth at both the author level and organizational level.

However, we did have limitations in our study. One of the limitations was that we only evaluated the field of AI. For a better understanding of the potential of this method, we need to test and evaluate this method in various fields. Another limitation was that we used some predefined metrics and computational operations to detect emergence. One might claim that other metrics and operations can work better. Using machine learning or deep learning methods might help to overcome this limitation in future works. Aslo, using our proposed attributes and metrics as the input and some emergent topics with the assigned emergence score by the expert as output can have a more automated and accurate way to identify emergent topics.